\documentclass{iopart}
\usepackage{epsfig}

\begin{document}

\letter{NMR characterization of spin-$\frac{1}{2}$ alternating
antiferromagnetic chains in the high-pressure phase of
(VO)$_{2}$P$_{2}$O$_{7}$}

\author{J Kikuchi\dag\footnote[3]{Present address: Institute for Solid
State Physics, University of Tokyo, Kashiwa, Chiba 277-8581, Japan
(kikuchi@issp.u-tokyo.ac.jp)}, K Motoya\dag, T Saito\ddag,
M Azuma\ddag\ and M Takano\ddag}

\address{\dag\ Department of Physics, Faculty of Science and 
Technology, Tokyo University of Science, Noda, Chiba 278-8510, Japan}

\address{\ddag\ Institute for Chemical Research, Kyoto University, Uji, 
Kyoto-fu 611-0011, Japan}

\begin{abstract}
\\    Local-susceptibility measurements via the NMR shifts of $^{31}$P and
    $^{51}$V nuclei in the high-pressure phase of (VO)$_{2}$P$_{2}$O$_{7}$
    confirmed the existence of a unique alternating antiferromagnetic chain
    with a zero-field spin gap of 34 K. The $^{31}$P nuclear spin-lattice
    relaxation rate scales with the uniform spin susceptibility below
    about 15 K which shows that the temperature dependence of both the
    static and dynamical spin susceptibilities becomes identical at
    temperatures not far below the spin-gap energy.

\end{abstract}





\nosections
Magnetic excitations of a low-dimensional quantum antiferromagnet have
been one of the current topics among the condensed matter physicists.
Vanadyl pyrophosphate (VO)$_{2}$P$_{2}$O$_{7}$ had long been believed as a
prototype of a spin-$\frac{1}{2}$ two-leg ladder which has a magnetic lattice
intermediate between one and two spatial dimensions
\cite{johnston87,barns94,eccleston94,troyer94,dagotto96}.  The ladder
model, however, has been rejected by an observation of a dominant
magnetic interaction perpendicular to the supposed ladder axis via the
inelastic neutron scattering (INS) measurements \cite{garrett97}.  A
dimerized (alternating) chain model has now been becoming accepted as
an alternative starting point,
although a mechanism of the major exchange interaction between distant pairs of
V$^{4+}$ spins via PO$_{4}$ tetrahedra is still under study
\cite{koo00,daku01,petit02}.

The INS experiments has also revealed the existence of the mode with a
gap nearly twice the gap of the lowest excited triplet which cannot be
explained by a simple alternating-chain model.  This mode has first
been assigned as a bound state of two magnons possibly formed via
interchain couplings \cite{uhrig98}, but it was difficult to account
for the intensity comparable to the fundamental mode.  Recent NMR
\cite{kikuchi99} and high-field magnetization \cite{yamauchi99}
studies have suggested on this issue that the two
structurally-distinguishable chains of V atoms, which were thought to
be magnetically identical, have different spin-gap energies.  This
gives a natural explanation for the existence of two distinct modes
with almost equal spectral weight, and has been supported by the
subsequent Raman-scattering experiments \cite{kuhlmann00} and
theoretical studies on relevant exchange interactions
\cite{koo00,daku01,petit02}.

The above confusion concerning the modelling and interpretation of the
experimental results of (VO)$_{2}$P$_{2}$O$_{7}$ comes not only from the
unexpectedly strong V-V exchange via PO$_{4}$ tetrahedra, but
also from the presence of structurally-inequivalent V chains
\cite{nguyen95,hiroi99}.  More recently, Azuma \etal have found that
(VO)$_{2}$P$_{2}$O$_{7}$ transforms into another phase with different
symmetry under pressure \cite{azuma99}.  All the V atoms occupies a
unique crystallographic site in the high-pressure (HP) phase, so that
the magnetic chains made of V$^{4+}$ spins are all equivalent.
Therefore, HP-(VO)$_{2}$P$_{2}$O$_{7}$ will be a better example of the
alternating antiferromagnetic chain with quantum spin $\frac{1}{2}$.
In this letter, we report microscopic characterization of the magnetic
chains in the HP phase of (VO)$_{2}$P$_{2}$O$_{7}$ via NMR. A single
spin component characterized by a zero-field gap of 34 K was found,
presenting support for the double-chain scenario for the ambient-pressure
(AP) phase.

Single crystals of the HP phase of (VO)$_{2}$P$_{2}$O$_{7}$ were grown
as described in \cite{saito00}.  Since the crystals were too
small to observe an NMR signal, they were crushed into powders and the
NMR measurements were made on these powders.  Standard spin-echo pulse
techniques were utilized for most of the experiments.
%
\begin{figure}
\begin{center}
\epsfxsize=70mm \epsfbox{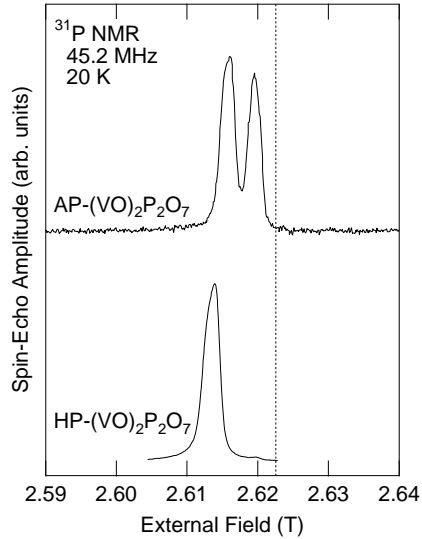}
\end{center}
\vspace{-5mm}
\caption{$^{31}$P NMR spectrum in the ambient-pressure 
(upper panel and the high-pressure (lower panel) phases of
(VO)$_{2}$P$_{2}$O$_{7}$ at 20 K. The dotted line indicates the
zero-shift position for $^{31}$P.}
\label{fig:31Pspectrum}
\end{figure}
\begin{figure}
\begin{center}
\epsfxsize=80mm \epsfbox{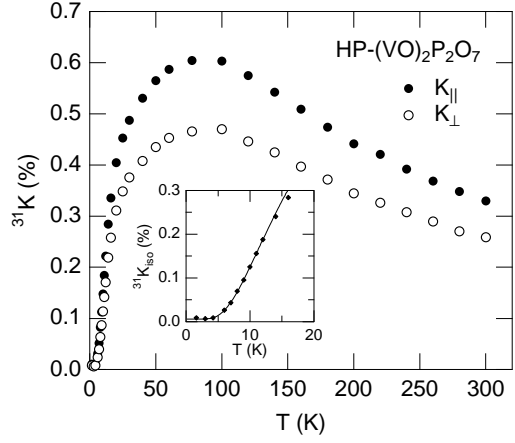}
\end{center}
\vspace{-5mm}
\caption{Temperature dependence of the principal values $K_{\|}$
and $K_{\bot}$ of the $^{31}$P NMR-shift tensor in the high-pressure
phase of (VO)$_{2}$P$_{2}$O$_{7}$.  The inset shows the isotropic
component of the NMR shift at low temperatures with the result of the
fitting (solid line).}
\label{fig:31KvsT}
\end{figure}
\begin{figure}
\begin{center}
\epsfxsize=75mm \epsfbox{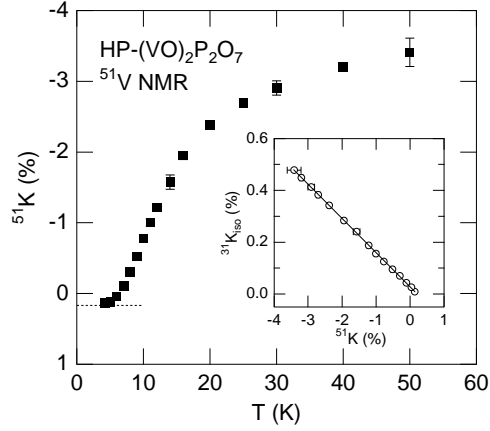}
\end{center}
\vspace{-5mm}
\caption{Temperature dependence of the $^{51}$V NMR shift in the
high-pressure phase of (VO)$_{2}$P$_{2}$O$_{7}$.  The dotted line
represents the orbital shift. The inset is a scaling of $^{51}K$ and
$^{31}K_{\rm iso}$ plotted with temperature the implicit parameter.}
\label{fig:51KvsT}
\end{figure}

An example of the field-swept $^{31}$P NMR spectrum in the 
HP phase of (VO)$_{2}$P$_{2}$O$_{7}$ is shown in
figure~\ref{fig:31Pspectrum}.  The spectrum in the AP phase \cite{kikuchi99} is
also shown for comparison.  The spectrum in the HP phase consists of a
single line as expected from the unique crystallographic site of
phosphor in the
unit cell. 
This is contrasted with the AP phase where the spectrum splits into two
groups of lines owing to the presence of two kinds of V chains with
different gap energies \cite{kikuchi99}.  The line-shape analysis
revealed that the symmetry of an NMR-shift tensor at the P site is
almost uniaxial.
Assuming the exact uniaxial symmetry, we determined the two independent
principal values $K_{\|}$ and $K_{\bot}$ corresponding to the shift
with the external field parallel and perpendicular to the local
symmetry axis, respectively.  The results are shown in
figure~\ref{fig:31KvsT} as a function of temperature.  Both $K_{\|}$
and $K_{\bot}$ scale the bulk magnetic susceptibility $\chi$ which is corrected
by subtracting the contribution of paramagnetic impurities.  Following
the standard $K$\textendash $\chi$ analysis, the tensor components of
the hyperfine coupling at the P site were determined as $A_{\|}=2.19$
T/$\mu_{\rm B}$ and $A_{\bot}=1.70$ T/$\mu_{\rm B}$.  These values
yield the isotropic and uniaxial components, $A_{\rm iso}=1.92$
T/$\mu_{\rm B}$ and $A_{\rm ax}=0.13$ T/$\mu_{\rm B}$, respectively.
$A_{\rm ax}$ is larger than and different in sign from that due to the
classical dipolar field of V$^{4+}$ spins $A_{\rm ax}^{\rm
dip}=-0.036$ T/$\mu_{\rm B}$, indicating that the V$^{4+}$ spins are
transferred not only to the P-$3p$ orbitals but also to the P-$3s$ orbital.

The susceptibility of a one-dimensional (1D) gapped spin system at
temperatures well below the gap $\Delta$ is proportional to
$T^{-1/2}\exp(-\Delta/T)$ \cite{troyer94}.  In order to determine $\Delta$, we
fitted the $T$ dependence of the isotropic component of the NMR shift
$^{31}K_{\rm iso}$ below 10 K to the form $^{31}K_{\rm
iso}=K_{0}+\alpha T^{-1/2}\exp(-\Delta(H)/T)$, where the reduction of
$\Delta$ by fields is explicitly written.  The result is shown in the
inset of figure~\ref{fig:31KvsT}.  The obtained parameters are
$K_{0}=0.006$ \%, $\alpha=0.081$ K$^{1/2}$, and $\Delta (2.62~\rm{T})$
= 31 K which gives $\Delta (0)$ = 34 K with the use of the measured
$g$ factor \cite{saito00}.  $\Delta(0)$ is in good agreement with that
evaluated from the bulk $\chi$ but is larger than the values
determined from the critical field of the magnetization process
($\sim$23 K) \cite{azuma99} and the INS on polycrystals ($\sim$25 K)
\cite{saito01} for unknown reasons.

A free-induction-decay (FID) signal of $^{51}$V 
has also been observed below about 50 K.
The spectrum was obtained by integrating the FID signal while sweeping
the external field.  The $T$ dependence of the $^{51}$V NMR shift
$^{51}K$ determined from the peak position of the spectrum is shown in
figure~\ref{fig:51KvsT}.  Also shown in the inset is a plot of
$^{31}K_{\rm iso}$ versus $^{51}K$ with $T$ the implicit parameter.  A
linear relation found between $^{31}K_{\rm iso}$ and $^{51}K$
demonstrates that the $T$ dependence of the local spin susceptibility
is identical for both the sites.  This is a clear sign of
HP-(VO)$_{2}$P$_{2}$O$_{7}$ having only one independent spin component.
The $T$ dependence of $^{51}K$ was analyzed in the same way as that of
$^{31}K_{\rm iso}$ using $\Delta$ determined above. The $T$-independent
orbital (van-Vleck) shift was then obtained to be 0.182 \%.  The hyperfine
coupling constant at the V site determined from the slope of the
$^{51}K {\textrm -}\chi$ plot is $-14.8$ T/$\mu_{\rm B}$, which is in a
reasonable range as a core-polarization field of a $3d$
transition-metal ion \cite{abragam70}.
\begin{figure}
\begin{center}
\epsfxsize=75mm \epsfbox{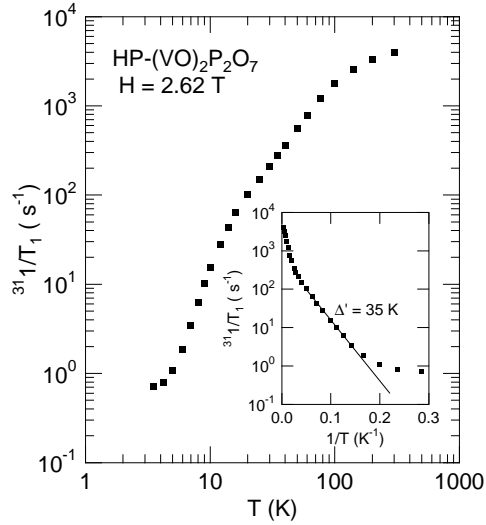}
\end{center}
\vspace{-5mm}
\caption{Temperature dependence of the $^{31}$P nuclear spin-lattice 
relaxation rate $1/T_{1}$ in the high-pressure phase of
(VO)$_{2}$P$_{2}$O$_{7}$.  The inset is a semi-logarithmic plot of
$1/T_{1}$ as a function of inverse temperature.  The solid line in the
inset shows the activation law $1/T_{1}\propto \exp(-\Delta^{\prime}/T)$
with $\Delta^{\prime}=35$ K.}
\label{fig:31WvsT}
\end{figure}

Figure~\ref{fig:31WvsT} shows the $T$ dependence of the $^{31}$P nuclear
spin-lattice relaxation rate $1/T_{1}$.  $T_{1}$ above 8 K was
determined as the time constant of the exponential recovery of $^{31}$P
magnetization $M(t)$.  Below 8 K where non-exponential recovery
appears, we analyzed $M(t)$ by fitting to the form
$1-M(t)/M(\infty)\propto \exp(-t/T_{1}-(t/\tau_{1})^{1/2})$ which
incorporates the relaxation rate $1/\tau_{1}$ due to paramagnetic
impurities \cite{mchenry72}.  As shown in the inset of
figure~\ref{fig:31WvsT}, $1/T_{1}$ exhibits activated behavior below
about 20 K. The exponential decrease of $1/T_{1}$ is, however, masked
below $\sim$8 K synchronizing the appearance of non-exponential recovery.
The asymptotic value of $1/T_{1}$ at low $T$ is suppressed by applying fields
as expected for the impurity-limited relaxation rate.
$1/T_{1}$ depends on $H$ at higher temperatures as well where the recovery
is exponential, but the $H$ dependence roughly follows the 1D
diffusive form $1/T_{1}\propto H^{-1/2}$ as observed in
AP-(VO)$_{2}$P$_{2}$O$_{7}$ \cite{kikuchi97}.  Details of the $H$
dependence of $1/T_{1}$ will be presented in a separated paper.
The activation energy was estimated as $\Delta^{\prime}=35$ K by fitting
the data between 8 and 20 K to the form $1/T_{1}\propto
\exp(-\Delta^{\prime}/T)$.  As the interbranch ($\Delta S_{z}=\pm 1$)
transitions within the lowest excited triplet \cite{sagi96} are expected
to dominate the nuclear-spin relaxation due to the
predominantly-isotropic hyperfine fields, the obtained
$\Delta^{\prime} $ would give an estimate of the zero-field gap. $\Delta^{\prime} $
indeed agrees well with $\Delta (0)$ evaluated from the NMR shift.
%
%
\begin{figure}
\begin{center}
\epsfxsize=75mm \epsfbox{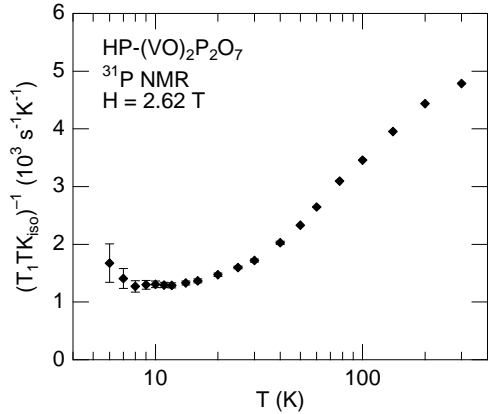}
\end{center}
\vspace{-5mm}
\caption{Temperature dependence of $(T_{1}TK_{\rm iso})^{-1}$ in the
high-pressure phase of (VO)$_{2}$P$_{2}$O$_{7}$.}
\label{fig:T1TKvsT}
\end{figure}

Figure~\ref{fig:T1TKvsT} shows the $T$ dependence of $1/T_{1}T$ divided
by $K_{\rm iso}$.  One of the remarkable features of the result is that the
ratio $(T_{1}T)^{-1}/K_{\rm iso}$ becomes $T$ independent below about
15 K. (An upturn below $\sim$7 K is due to the impurity contribution to
$1/T_{1}$ and is extrinsic.)  It is well known that, while the NMR
shift is proportional to the uniform static susceptibility
$\chi^{\prime}(0,0)$, $1/T_{1}$ samples the dissipative part of the
dynamical susceptibility $\chi^{\prime \prime}(q, \omega)$ at the
nuclear Larmor frequency $\omega_{n}$ \cite{moriya63};
\begin{eqnarray}
    \frac{1}{T_{1}}= \frac{k_{\rm B} T}{2 \mu_{\rm B}^{2}} 
    \sum_{q} \vert A(q)\vert^{2} \frac{\chi^{\prime \prime}(q,
    \omega_{n})}{\omega_{n}}.\nonumber
\label{MoriyaT1}
\end{eqnarray}

\noindent
Here $A(q)$ is the Fourier transform of the hyperfine coupling.  Since
$A(q)$ has a maximum at $q$ = 0, $1/T_{1}$ at the P site is most
sensitive to $\chi^{\prime \prime}(q\sim 0,\omega_{n})$ which is 
dominant at low $T$ in a gapped 1D spin system \cite{troyer94}.  The
$T$-independent behavior of $(T_{1}T)^{-1}/K_{\rm iso}$ therefore
indicates that the $T$ dependence of $\chi^{\prime \prime}(q\sim
0,\omega_{n})$ and $\chi^{\prime}(0,0)$ at low $T$ is identical and
should be described by a common energy gap.  Such a characteristic of
the magnetic excitations in a gapped 1D spin system has been predicted
theoretically based on a picture of free magnons \cite{troyer94}, but has
rarely been observed experimentally \cite{itoh97}.  To our knowledge,
this is the first experimental verification of $1/T_{1}T$ and $K$
having identical $T$ dependence at low $T$, not relying on any
model-dependent form of these quantities.  From the experimental
viewpoint, it is worth noting that the scaling between $1/T_{1}T$ and
$K$ holds below $T\sim \Delta/2$.  This suggests nearly free propagation
of magnons being realized at temperatures not far below $\Delta$.  It is
therefore practical to use experimental data in the region $T\leq
\Delta/2$ for a reliable estimate of $\Delta$, although the activated
behavior of physical quantities such as $\chi$ and $1/T_{1}$ is
theoretically justified only for $T\ll \Delta$ \cite{troyer94}.

Above about 20 K, the scaling breaks down and $(T_{1}TK_{\rm iso})^{-1}$
increases gradually with $T$.  This means that $\chi^{\prime
\prime}(q\sim 0,\omega_{n})$ grows more rapidly than
$\chi^{\prime}(0,0)$.  As the temperature is now comparable with or higher
than $\Delta$, interactions between magnons and/or the $q\neq 0$ component
of spin fluctuations will become increasingly important and would enhance
$\chi^{\prime \prime}(q,\omega)$ over $\chi^{\prime}(0,0)$.

In conclusion, we have measured $^{31}$P and $^{51}$V NMR in the
high-pressure phase of (VO)$_{2}$P$_{2}$O$_{7}$.  It was found that the
temperature dependence of the local static spin susceptibility at the
P site is identical with that at the V site.  The dynamical spin
susceptibility $\chi^{\prime \prime}(q,\omega)$ near $q=0$ also scales with
the static susceptibility at low temperatures below about a half of
the spin-gap energy which was estimated to be 34 K at zero field.  All of these
observations provides microscopic evidence for a unique kind of
magnetic chain existing in the high-pressure phase of
(VO)$_{2}$P$_{2}$O$_{7}$, as well as for coexistence of
magnetically-inequivalent chains in its ambient-pressure phase.

\end{document}